\documentclass[aps,prb,twocolumn,showpacs]{revtex4}
\usepackage{bm}
\usepackage{amssymb}
\usepackage{graphicx,epsfig}

\newcommand{\nn}{\nonumber \\}

\def\da     {\downarrow}
\def\ua     {\uparrow}
\def\rg     {\rangle}

\begin{document}
\title{ Commensurate quantum oscillations  in coupled qubits}
\author{Mun Dae Kim}
\email{mdkim@yonsei.ac.kr}
\affiliation{Institute of Physics and Applied Physics, Yonsei University, Seoul 120-749, Korea \\
Korea Institute for Advanced Study, Seoul 130-722, Korea}
\date{\today}

\begin{abstract}
We study the coupled-qubit oscillation driven by an oscillating field.
When the period of the non-resonant mode is commensurate with
that of the resonant mode of the Rabi oscillation, we show that
the controlled-NOT (CNOT) gate operation can be demonstrated.
For a weak coupling the CNOT gate operation is achievable by
the commensurate oscillations, while for a sufficiently strong coupling
it can be done for arbitrary parameter values.
By finely tuning the amplitude of oscillating field it is shown that the high fidelity of
the CNOT gate can be obtained for any fixed coupling strength and qubit energy gap in experiments.
\end{abstract}

\maketitle

\section{Introduction}

The universal gate for quantum computing consists of  the single
qubit and the entangling two qubit operations. Usually,
electromagnetic oscillating fields such as microwave fields, laser
pulses, and oscillating voltages have been used for single qubit
operations. The atomic states of cavity-QED \cite{Raimond} and
ion-trap \cite{Cirac} qubits are used as a natural basis of qubit
states. In these cases the quantum Rabi-type oscillation can be
analyzed in an approximation, called the rotating wave
approximation (RWA). Similarly, the semi-classical Rabi-type
oscillation of qubits of the artificial atomic states such as the
superconducting charge qubit employed in the circuit-QED quantum
computing \cite{Blais} and the flux qubit \cite{Mooij} can also be
analyzed in this approximation.
%


We study the coherent two-qubit oscillation driven by
an oscillating field. The two-qubit oscillation enables the two-qubit gate for
the quantum computing. Among the two-qubit gates the CNOT gate is the most basic
two-qubit operation for the universal gate. \cite{Barenco}
The CNOT gate operation was achieved in superconducting qubits
with Ising-type interaction without driving oscillating field. \cite{Yamamoto}
The oscillating-field-driven CNOT gate operation
in superconducting qubits has recently been
reported, but the fidelity is not so high due to the weak coupling strength
between qubits. \cite{Plantenberg}
In this study we propose a scheme for the CNOT gate operation
between coupled qubits  under an oscillating field
for general Ising-type coupling strength.
We discuss the CNOT gate operation for both
the strong and weak coupling strength and
show that the high fidelity of CNOT gate can be obtained even for
a weak coupling.


The CNOT gate uses the discriminating operations corresponding to
different states of control qubit. Depending on the control qubit
states, the coupled-qubit state demonstrates a Rabi-type
oscillation for the resonant oscillating field
or non-resonant oscillation. During the
$\pi$ rotation  the target qubit state flips to the other qubit
state for a control qubit state, while for the other control qubit
state the target qubit state goes back to the original state
before reaching the other state. Though  the latter oscillation is
far from the Rabi-type oscillation, the  {\it commensurate
oscillations} give rise to the CNOT gate operation.
By using the  commensurate mode oscillations
we obtain high fidelity for the CNOT gate operation.
We show that for a weak coupling
a high performance CNOT gate can be achieved by tuning the parameters,
while for a sufficiently strong coupling the maximum fidelity can be
obtained regardless of the values of system parameters.
The maximum fidelity for a weak coupling can be obtained
for any fixed coupling strength and qubit energy gap by finely tuning
the amplitude of oscillating field in an experimental setup.
This scheme of using the commensurate oscillation is quite general,
so it is applicable to the natural atomic qubits as well as the
solid state qubits.

\section{Hamiltonian of coupled qubits}

The Hamiltonian of a two level system (qubit {$a$) driven by an
oscillating field with frequency $\omega$ is given by
\begin{eqnarray}
H^a=- E^a_z(\kappa,t)\sigma_z- t^a_q\sigma_x,
\end{eqnarray}
where
\begin{eqnarray}
E^a_z(\kappa,t)=\frac{E^a(\kappa)}{2}+g\cos\omega t,
\end{eqnarray}
$\kappa$ is the external variable controlling the qubit energy levels,
and $\sigma_{z,x}$ are the Pauli matrices.
Here g is the coupling strength between the qubit and the oscillating field
which is proportional to the amplitude of the oscillating field,
$t_{\rm q}$ is the tunnelling amplitude between different (pseudo-) spin states,
and $E^a(\kappa)=E^a_{\uparrow}(\kappa)-E^a_{\downarrow }(\kappa)$.
The qubit energy gap $E^a(\kappa)$ can be controlled by $\kappa$, and
for a particular value of $\kappa_0$ the qubit can be brought to the degeneracy point,
$E^a_{\uparrow}(\kappa_0)=E^a_{\downarrow }(\kappa_0)\equiv E^a_0$.

At this point the dominant energy scale is the tunnelling energy
$t^a_q$ and, if we introduce the coordinate transformation,
 $|0\rangle=(|\da\rangle+|\ua\rangle)/\sqrt{2}$
and $|1\rangle=(|\da\rangle-|\ua\rangle)/\sqrt{2}$,
we have the Hamiltonian:
\begin{eqnarray}
{\cal H}^a\!\!\!&=&\!\!\!{\cal E}^a_{0}|0\rangle\langle 0|\!+\!{\cal E}^a_{1}|1\rangle\langle 1|
\!+\!g \cos\omega t (|1\rangle\langle 0|\!+\!|0\rangle\langle 1|),
\label{OneT}
\end{eqnarray}
where ${\cal E}^a_{0(1)} = E^a_0\mp t^a_q$.
Hence the resonant microwave with frequency $\omega=2t^a_q$
gives rise to the Rabi-type oscillation between the qubit states,
$|0\rangle$ and $|1\rangle$.

When two qubits (qubits $a$ and $b$) are coupled, the  Hamiltonian for coupled qubits
in the basis of $\{|\da\rangle, |\ua\rangle\}$ can be written as
\begin{eqnarray}
H&=&H^a\otimes I+I\otimes H^b-t^a_q \sigma_x\otimes I-t^b_qI\otimes \sigma_x\nn
&&+J\sigma_z\otimes\sigma_z
\end{eqnarray}
with  the Ising-type coupling strength  $J$  ~\cite{KimHong,KimTunable}
\begin{eqnarray}
J=\frac14(E_{\downarrow\downarrow}+E_{\uparrow\uparrow}-E_{\downarrow\uparrow}
-E_{\uparrow\downarrow}).
\label{J}
\end{eqnarray}
Here, $E_{ss'}$ are the energy levels of coupled qubits,
$E_{ss'}(\kappa_a,\kappa_b)=E^a_s(\kappa_a)+E^b_{s'}(\kappa_b)\pm J$
($+$ for $s=s'$, $-$ for $s=-s'$),
where $\kappa_a$ and $\kappa_b$ are the control variables for qubit $a$ and $b$, respectively.
We here neglect the two-qubit tunneling term, $t^{ab}_q|s,s'\rangle\langle -s,-s'|$,
because it is negligibly small for usual parameter regimes.
This term gives rise to the XY-type interaction which enables
the SWAP gate operation rather than the CNOT gate.


We can rewrite the Hamiltonian  as
\begin{eqnarray}
\label{TwoH}
H \!\!\! &=& \!\!\! \sum_{s,s'}F_{ss'}(\kappa_a,\kappa_b,t)|s,s'\rangle\langle s,s'|
 - t^a_{q}|s,s'\rangle\langle -s,s'| \nonumber\\
&&- t^b_{q}|s,s'\rangle\langle s,-s'|,
\end{eqnarray}
where $-s$ is the opposite spin of $s$, and
\begin{eqnarray}
{F}_{{\downarrow\downarrow}}(\kappa_a,\kappa_b,t)   &=&   E_{{\downarrow\downarrow}}(\kappa_a,\kappa_b)-2g  \cos\omega t, ~~\nonumber\\
~{F}_{{\downarrow\uparrow}}(\kappa_a,\kappa_b,t)   &=&   {E}_{{\downarrow\uparrow}}(\kappa_a,\kappa_b),\\
~{F}_{{\uparrow\uparrow}}(\kappa_a,\kappa_b,t)   &=&  {E}_{{\uparrow\uparrow}}(\kappa_a,\kappa_b)+2g  \cos\omega t, ~~~\nonumber\\
{F}_{{\uparrow\downarrow}}(\kappa_a,\kappa_b,t)   &=&   {E}_{{\uparrow\downarrow}}(\kappa_a,\kappa_b),\nonumber
\end{eqnarray}
Afterward, we will omit $\kappa_a,\kappa_b$ in $E_{ss'}(\kappa_a,\kappa_b)$ for simplicity.

\begin{figure}[b]
\hspace{-0.5cm}
\includegraphics[height=5cm]{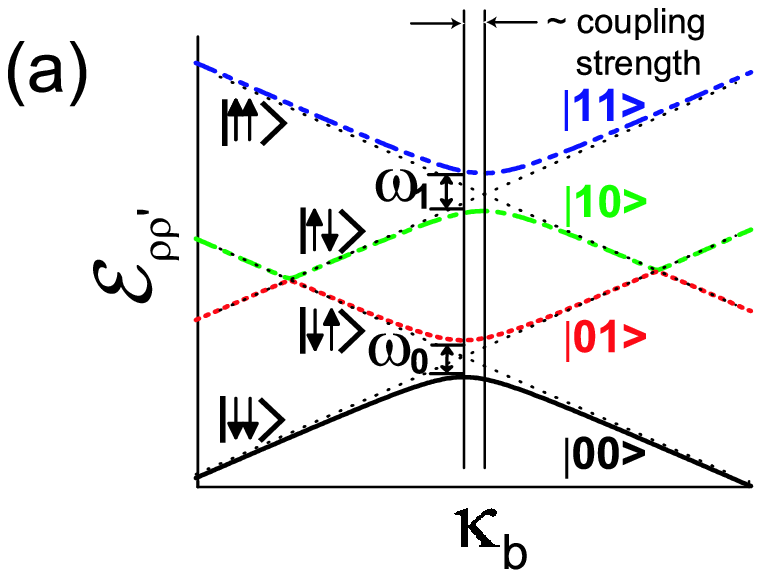}
\hspace{0.8cm}
\includegraphics[height=4.5cm]{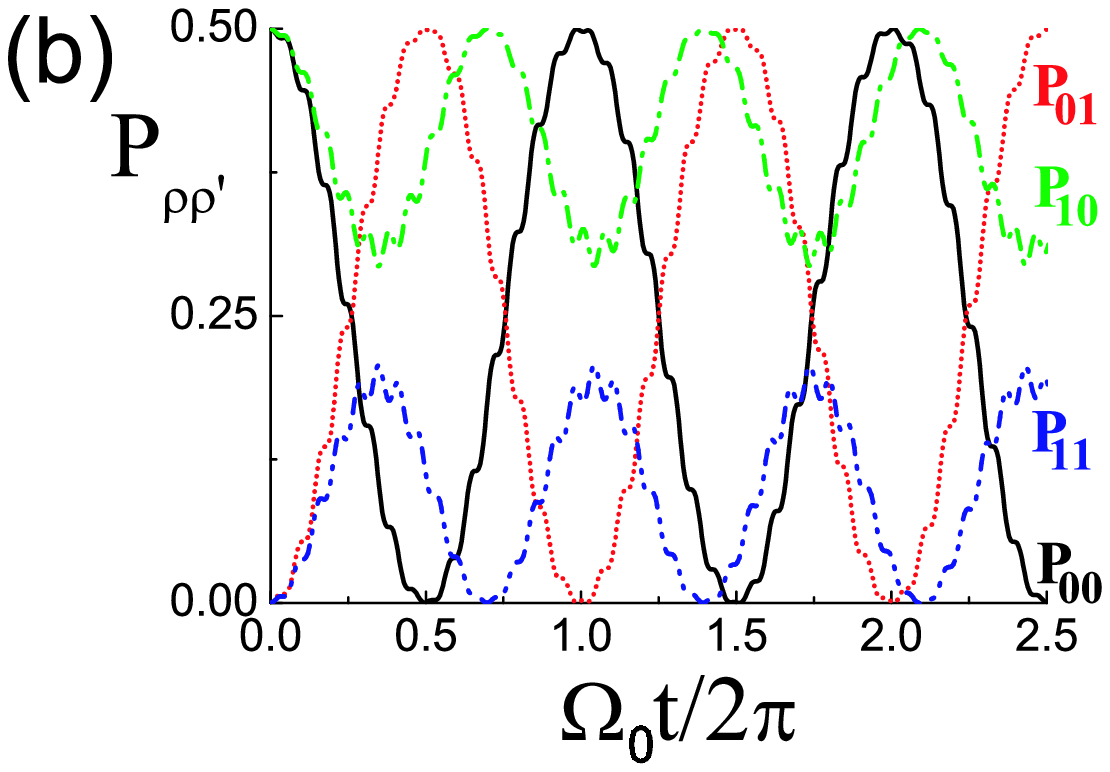}
\vspace{-0.2cm}
%
%
\caption{(Color online) (a) Energy levels ${\cal E}_{\rho\rho'}$ of coupled
qubits, where $\rho, \rho' \in \{0, 1\}$.
$E_{ss'}$ with $s,s' \in \{\da,\ua\}$ are shown as thin dotted lines.
The distance between two degeneracy points corresponds to the coupling strength between
two qubits.
(b) Occupation probabilities of $|\rho\rho'\rangle$
states during Rabi-type oscillations
at the lower degeneracy point where $E_{\da\da}=E_{\da\ua}$.
Here we use the parameter values such that coupling strength $J/h=$0.6GHz,
qubit energy gap $\omega_0/2\pi$=4GHz, and Rabi frequency $\Omega_0/2\pi=600$MHz.
The initial state is chosen as
$\psi(0)=(|00\rangle+|10\rangle)/\sqrt{2}$ and the CNOT gate is expected to
be achieved at $\Omega t=$ (odd) $\pi$.} \label{level}
\label{weak}
\end{figure}

To perform a CNOT gate operation, the system parameters should be
adjusted. We consider the situation that the external variables are adjusted
in a way that (i) the energy gap  $\Delta E=E_{ss'}-E_{-ss'}$
 between different control qubit (qubit $a$) states
is large and (ii) the target qubit (qubit $b$)  is at a degeneracy point, $E_{\da\da}=E_{\da\ua}$.
This situation is usual for the CNOT gate operation of coupled qubits. \cite{Yamamoto,KimHong}
In this case the tunneling process $t^b_{q}$ between different target qubit states takes a role,
while $t^a_{q}$  between different control qubit states is negligible
due to the large energy level difference $\Delta E$.
Consequently, the two-qubit Hamiltonian $H$ becomes
block-diagonal.

Figure \ref{weak}(a) shows the energy levels $E_{ss'}$
as a function of $\kappa_b$, where we choose $\kappa_a$ such that $|E_{ss'}-E_{-ss'}|\gg t^a_q$
and thus $t^a_q$ can be negligible.
In the figure there are two degeneracy points;
lower degeneracy point where $E_{\da\da}=E_{\da\ua}$ and upper degeneracy point where $E_{\ua\da}=E_{\ua\ua}$.
By adjusting the variable $\kappa_b$, the coupled-qubit states can be
brought to one of these degeneracy points. Here the distance between these degeneracy points
is related to the coupling strength  between two qubits. \cite{Yamamoto,KimHong}

Here we introduce a coordinate  transformation,
\begin{eqnarray}
V=\exp\left(-\frac{i}{2}\sigma_y \theta_\da\right) \!\!\oplus
\exp\left(-\frac{i}{2}\sigma_y \theta_\ua\right)
\end{eqnarray}
with $\tan\theta_s =2t^b_{q}/|E_{s\downarrow}-E_{s\uparrow}|$
in order to couple the oscillating field with the off-diagonal
elements of the qubit Hamiltonian.
Then the Hamiltonian ${\cal H}=V^{-1}HV$ becomes
\begin{eqnarray}
{\cal H}&=&\sum_{{\rho}=0,1} \Big[{\cal E}_{{\rho}0}(t)|{\rho}0\rangle\langle {\rho}0| +{\cal E}_{{\rho}1}(t)|\rho 1\rangle\langle {\rho}1|\nn
&& + \alpha_\rho g \cos\omega t (|\rho 0\rangle\langle \rho 1|+|\rho 1\rangle\langle \rho 0|)\Big],
\label{TwoT}
\end{eqnarray}
where
\begin{eqnarray}
{\cal E}_{{\rho\rho'}}(t)  = {\cal E}^0_{{\rho\rho'}}
-[(-1)^\rho +(-1)^{\rho'}\!\beta_\rho] g \cos\omega t
\end{eqnarray}
with
$\alpha_0=\sin\theta_\da, ~\beta_0=\cos\theta_\da, ~\alpha_1=\sin\theta_\ua$
and $\beta_1=\cos\theta_\ua$.
The two-qubit states $|\rho\rho'\rangle$ are given as
\begin{eqnarray}
|\rho 0\rg &=&\cos(\theta_s/2)|s\da\rg +\sin (\theta_s/2)|s\ua\rg,\nonumber\\
|\rho 1\rg &=&-\sin(\theta_s/2)|s\da\rg +\cos(\theta_s/2)|s\ua\rg,
\label{transform}
\end{eqnarray}
where $s=\da(\ua)$ for $\rho=0(1)$.

At the lower degeneracy point $E_{\da\da}=E_{\da\ua}\equiv E_0$, we have the relations,
\begin{eqnarray}
{\cal E}^0_{00}&=& -t^b_{q} +E_{0},~~~
{\cal E}^0_{01}=t^b_{q} +E_{0},\\
{\cal E}^0_{10(11)}&=&\frac{E_{\ua\ua}+E_{\ua\da}}{2}
\mp \sqrt{\left(\frac{E_{\ua\ua}-E_{\ua\da}}{2}\right)^2+(t^b_{q})^2},\nonumber
\end{eqnarray}
and
\begin{eqnarray}
\label{tan}
\theta_\da=\pi/2 ~~~{\rm and}~~~ \theta_\ua= \tan^{-1} (t^b_{q}/2J)
\end{eqnarray}
by using the relation of Eq. (\ref{J}).
%
Since the Hamiltonian ${\cal H}$ is block-diagonal,
we have
\begin{eqnarray}
{\cal H}&=&{\cal H}_0\oplus {\cal H}_1,\\
\label{H0}
{\cal H}_0&=&-\frac{\hbar \omega_0}{2}(|{0}0\rangle\langle {0}0| -|01\rangle\langle {0}1|) \nonumber\\
&&+  g \cos\omega t (|00\rangle\langle 01|+|01\rangle\langle 00|),\\
\label{H1}
{\cal H}_1 &=& -\frac{\hbar\omega_1+2 \beta_1 g\cos\omega t}{2}(|{1}0\rangle\langle {1}0| -|11\rangle\langle {1}1|) \nonumber\\
  &&+ \alpha_1 g\cos\omega t (|10\rangle\langle 11|+|11\rangle\langle 10|),
\end{eqnarray}
where
\begin{eqnarray}
\alpha_1&=&\sin\theta_\ua=t^b_q/\sqrt{(2J)^2+(t^b_q)^2}, \\
\beta_1&=&\cos\theta_\ua=2J/\sqrt{(2J)^2+(t^b_q)^2}. \nonumber
\end{eqnarray}
Here, $\alpha_1 g$ corresponds to the transition frequency between
$|10\rangle$ and $|11\rangle$ states, while $\beta_1 g$ term
induces unnecessary complicate oscillations.
In Fig. \ref{weak} the energy gaps are given as
\begin{eqnarray}
\hbar\omega_0&=&{\cal E}^0_{01}-{\cal E}^0_{00}=2t^b_{q} \\
\hbar\omega_1&=&{\cal E}^0_{11}-{\cal E}^0_{10}=2\sqrt{\left[(E_{\ua\ua}-E_{\ua\da})/2\right]^2+(t^b_{q})^2},
\nonumber
\end{eqnarray}
where $\hbar\omega_0$ is the qubit energy gap, and
$\hbar\omega_1$
depends on the qubit coupling strength through the relation
\begin{eqnarray}
\omega^2_1 =\omega^2_0+\left(\frac{4J}{\hbar}\right)^2.
\label{omega1}
\end{eqnarray}

We here consider the case that  the oscillating field is resonant
with the energy gap $\omega_0$ between the states $|00\rangle$ and $|01\rangle$,
{\it i.e.} $\omega=\omega_0$, at the degeneracy point $E_{\da\da}=E_{\da\ua}$. 
Then this Hamiltonian describes the usual Rabi-type oscillation between
the states  $|00\rangle$ and $|01\rangle$ with the Rabi frequency $\Omega_{\rm R} \approx g/\hbar$,
while  the evolution of  the states $|10\rangle$ and $|11\rangle$
is far from the Rabi oscillation, since the energy level difference
is not resonant with the oscillating field  frequency, $\omega_1 \neq \omega$, for a finite coupling strength $J$.

We again introduce a rotating coordinate such as $\psi(t)=U(t)\phi(t)$, where
\begin{eqnarray}
U(t)\!=\exp\left[\frac{i}{2}\omega_0 t\sigma_z\right] \!\!\oplus
\exp\left[\frac{i}{2}\left(\omega_0 t+\frac{2g\beta_1}{\omega_0}\sin\omega_0 t\right)\sigma_z\right].\nonumber\\
\end{eqnarray}
Accordingly,  the Schr{\" o}dinger equation
${\cal H}\psi(t)=i\hbar \frac{\partial}{\partial t} \psi(t)$
is written as
$i \hbar\frac{\partial}{\partial t} \phi(t)=\tilde{{\cal H}} \phi (t)$
with $\tilde{{\cal H}}=U^{-1}(t){\cal H}U(t)-i\hbar U^{-1}(t)(dU(t)/dt)
=\tilde{{\cal H}}_0\oplus\tilde{{\cal H}}_1$, where
\begin{eqnarray}
\label{tilH0}
\tilde{{\cal H}}_0 &=&
\left(\matrix{ 0 & g  \cos\omega_0 t e^{-i\omega_0 t} \cr
g  \cos\omega_0 t e^{i\omega_0 t} & 0 }\right), \\
\label{tilH1}
\tilde{{\cal H}}_1 &=&
\left(\matrix{ \omega_1-\omega_0 & g\alpha_1  \cos\omega_0 t e^{-i\xi(t)} \cr
g\alpha_1  \cos\omega_0 t e^{i\xi(t)} & \omega_1-\omega_0}\right),
\end{eqnarray}
and $\xi(t)=\omega_0 t+2g\beta_1 \sin\omega_0 t/\omega_0$.
From this Hamiltonian we can obtain the two-qubit oscillation
numerically. Also the dynamics can be analyzed in the RWA.

\section{Rotating wave approximation}

The RWA assumes near resonance $\omega \approx \omega_0$ and  weak
coupling between a qubit and a oscillating field $g/\hbar \ll
\omega_0$. \cite{Jaynes} In the quantum Rabi oscillation for
cavity-QED and ion-trap qubit, usually  $g/\hbar\omega_0 \approx
10^{-6}-10^{-7}$. For the usual superconducting qubits the
coupling strength $g/\hbar\omega_0 \sim $ O($10^{-1}$)
~\cite{Blais,Mooij} which is relatively strong, but we find that
the RWA gives accurate results consistent with our numerical
calculation.

The off-diagonal element of $\tilde{{\cal H}}_{0(1)}$
is written as
\begin{eqnarray}
\left[\tilde{{\cal H}}_0\right]_{12} \!\!\!\!\!&=&\!\!\! \frac{g}{2}(1+ e^{-2i\omega_0 t}), \\
\left[\tilde{{\cal H}}_1\right]_{12} \!\!\!\!\!&=&\!\!\! \frac{g\alpha_1}{2}
\!\!\sum_n J_n\!\!\left(\!\frac{2g\beta_1}{\omega_0}\!\right)\!\!
\left[e^{-in\omega_0 t}\!\!+\!e^{-i(n+2)\omega_0 t}\right],
\end{eqnarray}
where $J_n(x)$ is the Bessel function of the first kind.
In the usual RWA, the fast oscillating term $e^{2i\omega_0 t}$ in $[\tilde{{\cal H}}_0]_{12}$ is neglected.
Similarly we here neglect $e^{in\omega_0 t}  ~(n\neq 0)$ in $[\tilde{{\cal H}}_1]_{12}$,  resulting in
\begin{eqnarray}
\tilde{{\cal H}}^{\rm RWA}_0\!\!\!&=&\!\!\!\left(\matrix{0 & g/2 \cr g/2 & 0}\right), \\
\tilde{{\cal H}}^{\rm RWA}_1\!\!\!&=&\!\!\!\left(\matrix{\hbar(\omega_1-\omega_0) & g'/2  \cr
g'/2 & \hbar(\omega_1-\omega_0)}\right),
\end{eqnarray}
where
$g'=g\alpha_1 \left[J_0\left(2g\beta_1/\hbar\omega_0\right)
+J_{-2}\left(2g\beta_1/\hbar\omega_0\right)\right].$

Hence the Hamiltonian $\tilde{{\cal H}}^{\rm RWA}_0=(1/2)g\sigma_x=(g/\hbar)S_x$ describes the Rabi oscillation with
the Rabi frequency
$\Omega_{\rm R}=\Omega_0=g/\hbar,$
while the Hamiltonian $\tilde{{\cal H}}^{\rm RWA}_1$ shows a non-resonant oscillation
with the oscillating frequency
$\Omega_1 = \sqrt{\left(\omega_1-\omega_0\right)^2+(g'/\hbar)^2}.$
From the relation of Eq. (\ref{omega1}) we see that the behavior of
this non-resonant oscillation depends on the coupling strength $J$ as well as $\omega_0$ and $g$.

In Fig. \ref{weak}(b) we show the resonant and non-resonant oscillations,
when $\kappa_b$ is adjusted to the lower degeneracy point where $E_{\da\da}=E_{\da\ua}$.
Then a microwave with resonant frequency $\omega=\omega_0$ gives rise to
the Rabi oscillation between two states $|00\rangle$ and $|01\rangle$,
while the states $|10\rangle$ and $|11\rangle$ experience a non-resonant oscillation.
The controlled-NOT gate operation requires that the target qubit flips for a specific state
of control qubit such that $|00\rangle \rightarrow |01\rangle$ while $|10\rangle \rightarrow |10\rangle$.
However, for example, at $\Omega t=\pi$ in Fig. \ref{weak}(b)
the states $|11\rangle$ and $|10\rangle$ also evolves during
the transition from $|00\rangle$  to $|01\rangle$. Thus we cannot
expect a good CNOT gate operation in this case.

Although, for $\omega_1$ different from the resonant value of $\omega=\omega_0$,
the oscillation is not a Rabi oscillation,
the oscillation period can be
an even integer multiple of that of the resonant Rabi oscillation mode
for a specific values for parameters $g$, $J$, and $\omega_0$
which correspond to the oscillating field amplitude, the coupling strength
between qubits, and the qubit energy gap, respectively.
The condition for this commensurate periods is given by
\begin{eqnarray}
2n\frac{2\pi}{\Omega_1}=\frac{2\pi}{\Omega_0},
\label{condition}
\end{eqnarray}
as we can see in Fig. \ref{TwoRabi}.
This condition determines the value of $g$
for given values of $(\omega_0, J)$.

\begin{figure}[b]
\includegraphics[height=5.5cm]{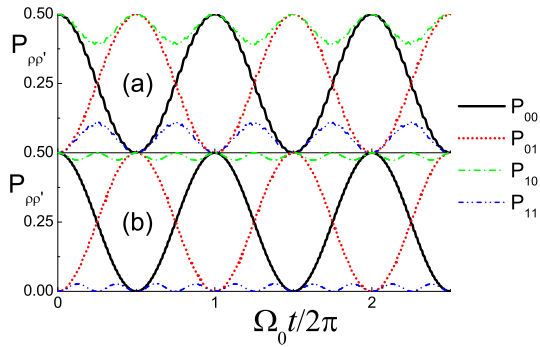}
\vspace*{-0.5cm}
\caption{(Color online) (a) Commensurate oscillations of occupation probability of coupled-qubit
states with the initial state, $|\psi(0)\rg=(|00\rg+|10\rg)/\sqrt{2}$ for $g/h=0.265$GHz. The non-resonant
oscillation modes ($P_{10}$ and $P_{11}$) are commensurate with
the resonant modes ($P_{00}$ and $P_{01}$). At $\Omega t=$
(odd)$\pi$, $P_{10}$ and $P_{11}$ recover their initial values,
thus the CNOT gate operation is achieved.
Here $\Omega_0= g/\hbar$, $J/h=0.5$GHz, and $\omega_0/2\pi=$4.0GHz.
(b) Higher order commensurate modes for smaller
$g/h=0.122$GHz with the same $J$ and $\omega_0$.
}
\label{TwoRabi}
\end{figure}

From Eq. (\ref{condition}) the value of $g$
for fidelity maxima can be expressed as
\begin{eqnarray}
\label{g}
g=\frac{\hbar(\omega_1-\omega_0)}{\sqrt{(2n)^2-\alpha^2_1\left[J_0\left(\frac{2g\beta_1}{\hbar\omega_0}\right)+
J_{-2}\left(\frac{2g\beta_1}{\hbar\omega_0}\right)\right]^2}}.
\end{eqnarray}
The argument of the Bessel function is written as
$2g\beta_1/\hbar\omega_0=(2g/\hbar\omega_0)4J/\sqrt{(\hbar\omega_0)^2+16J^2}$.
For $x\rightarrow 0$, the Bessel functions $J_0(x)$ and $J_{-2}(x)$ approach 1 and 0, respectively.
Thus, for small $J$ and large $\omega_0$ 
the expression of $g$ in Eq. (\ref{g}) can be approximated as
\begin{eqnarray}
g\approx \frac{1}{\sqrt{4n^2-\alpha^2_1}}\left(\sqrt{16J^2+(\hbar\omega_0)^2}-\hbar\omega_0\right),
\label{ga}
\end{eqnarray}
using  Eq. (\ref{omega1}).
These expressions of Eqs. (\ref{g}) and (\ref{ga})  provide the value for $g$ for the fidelity maxima
with  given values of $J$  and $\omega_0$.


\section{CNOT gate operation using commensurate modes}

The scheme for CNOT gate operation in this study uses the non-Rabi
oscillations for $|10\rg$ and $|11\rg$ states which are
commensurate with the Rabi oscillation for $|00\rg$ and $|01\rg$
states. In Fig. \ref{TwoRabi}  we display the numerical results
obtained from the Hamiltonian in Eqs. (\ref{tilH0}) and
(\ref{tilH1}), which show such  commensurate mode oscillations.
The initial state, $|\psi(0)\rg=(|00\rg+|10\rg)/\sqrt{2}$,
is driven by  an oscillating field with the resonant frequency $\omega=\omega_0 <\omega_1$.

\begin{figure}[t]
\includegraphics[height=5.5cm]{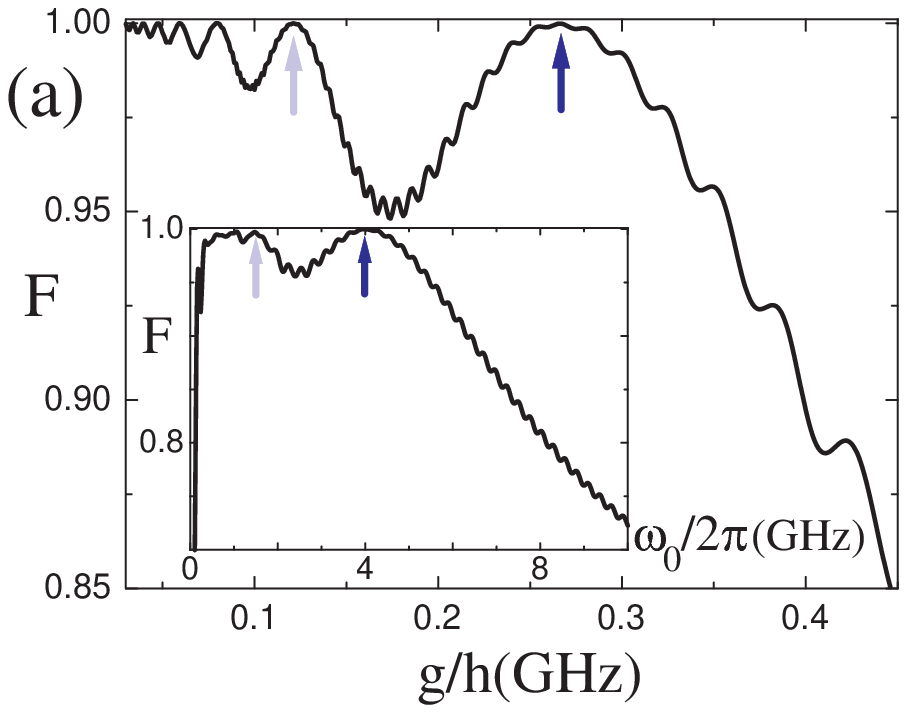}
\hspace{0cm}
\includegraphics[height=5.5cm]{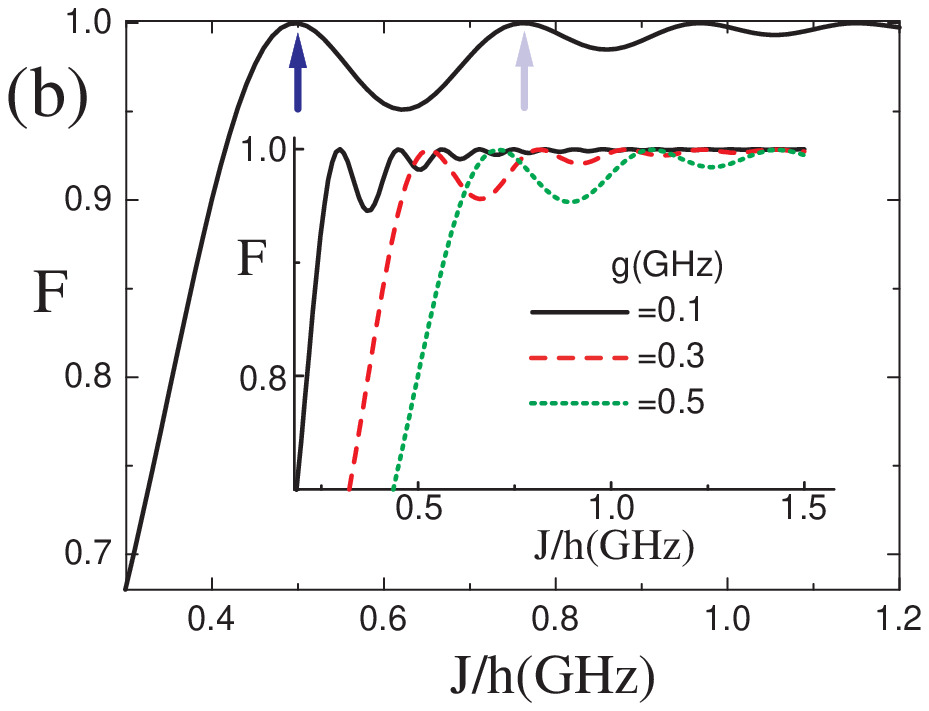}
%
\vspace*{-0.2cm}
\caption{ (Color online) (a) Fidelity F for CNOT gate
as a function of $g$ at $\Omega_0 t=\pi$. The main (dark arrow) and subsidiary (light arrow) maxima correspond to
the commensurate modes in Fig. \ref{TwoRabi}(a) and (b), respectively.
Here we set $J/h=0.5$GHz and $\omega_0=4$GHz.
Inset shows the fidelity envelops as a function of $\omega_0$
for $g/h=0.265$GHz and $J/h=0.5$GHz.
(b) Fidelity F as a function of $J$ for $g/h=0.265$GHz and $\omega_0=4$GHz.
Inset shows F for different $g$'s.}
\label{peak}
\end{figure}

In experimental situations
usually the coupling $J$ and the qubit energy gap $\omega_0$ are
set to be fixed. Thus the control of oscillating field amplitude
$g$ with fixed $J$ and $\omega_0$ will be more desirable.
For any given pair of $(\omega_0, J)$
one can find a commensurate oscillation  by finely tuning the value of $g$
according to Eq. (\ref{g}).
By varying  $g$ with fixed   $\omega_0$ and  $J$,
we were able to find a commensurate oscillation mode numerically; the oscillation period
of the  $|00\rg$ and $|01\rg$  states
is twice of that of the $|10\rg$ and $|11\rg$  states [Fig. \ref{TwoRabi}(a)],
which corresponds to $n=1$ in Eq. (\ref{condition}).
As $g$ decreases further,
another commensurate mode with a shorter period appears
in Fig. \ref{TwoRabi}(b) ($n=2$).
Actually we have found a series of commensurate modes as $g$ decreases.
The values of $g$ obtained numerically coincide well
with those from the RWA in Eq. (\ref{g}) as shown in Table \ref{table}.


The CNOT gate operation is done when the occupation probability
$P_{00}$ ($P_{01}$) is reversed perfectly from 0.5 (0) to 0 (0.5)
at $\Omega t=$ (odd) $\pi$. At the same time,
we can observe that the probabilities $P_{10}$ and $P_{11}$
recover their initial values 0.5 and 0, respectively.
As a result,  the CNOT operation is realized
by using these  commensurate oscillations.

\begin{figure}[t]
\vspace{0.8cm}
\includegraphics[height=5.5cm]{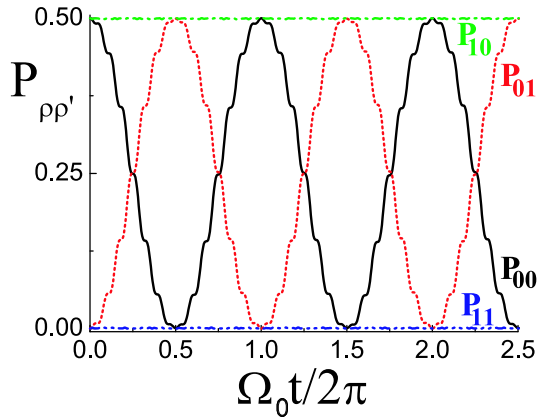}
\vspace{-0.2cm}
\caption{(Color online)  Rabi-type oscillations of occupation probabilities of $|\rho\rho'\rangle$
states for  strongly coupled qubits with the initial state $\psi(0)=(|00\rangle+|10\rangle)/\sqrt{2}$.
Here the parameters are $J/h=5$GHz,  $\omega_0/2\pi$=4GHz, and
$\Omega_0/2\pi=600$MHz at the degeneracy point where $E_{\da\da}=E_{\da\ua}$ in Fig. \ref{weak}(a).
}
\label{PC}
\end{figure}

Let us consider a concrete example for comprehensive understanding.
For superconducting flux qubits, \cite{Mooij,flux} $g=mB$ is the coupling between  the amplitude $B$
of the magnetic microwave field and the magnetic
moment $m$, induced by the circulating current, of the qubit loop.
In order to adjust the value of $g$, actually we need to vary the microwave
amplitude $B$, because the qubit magnetic moment is fixed at a specified degeneracy point.
The Rabi-type oscillation occurs between the transformed states
$|0\rangle=(|\da\rangle+|\ua\rangle)/\sqrt{2}$ and $|1\rangle=(|\da\rangle-|\ua\rangle)/\sqrt{2}$.
The states of qubits can be detected by shifting the magnetic pulse adiabatically
\cite{Kaku}. Since these qubit states are the superposition of the clockwise and counterclockwise
current states, $|\da\rangle$ and $|\ua\rangle$,
the averaged current of qubit states vanishes at the degeneracy point in Fig. \ref{weak}(a).
Thus, one can apply a finite dc magnetic pulse to
shift the qubits slightly away from the degeneracy point to detect
the qubit current states.

We now discuss the performance of CNOT gate operation. The
fidelity  for  CNOT gate operation is given by $F(t)={\rm
Tr}(M(t)M_{\rm CNOT})/4$,\cite{Plantenberg} where $M_{\rm CNOT}$ is the matrix for
the perfect CNOT operation and $M(t)$ is the truth table amplitude
at time $t$. In Fig. \ref{peak} we plot the  fidelity $F$ by
varying $g, ~\omega_0$ or $J$ at $\Omega_0 t=\pi$. In Fig.
\ref{peak}(a) the main and subsidiary maxima correspond to the CNOT
operation in Fig. \ref{TwoRabi}(a) and (b), respectively.
As shown in Eq. (\ref{g}) the fidelity maxima are determined by the three
parameters, $g, ~\omega_0$ or $J$. The series of maxima correspond to the different $n$ in Eq. ({\ref{g}).
In the inset of Fig. \ref{peak}(a) we also
show the fidelity as a function of $\omega_0$. 

An interesting behavior of fidelity maxima is shown in Fig. \ref{peak}(b)
as a function of $J$, where the  fidelity
approaches the maximum  as the coupling strength $J$ increases.
In the inset we show the oscillations for various parameters converge to 1, which
implies that for sufficiently strong coupling
maximum fidelity for the CNOT gate is achievable regardless of the values of other
parameters.
This is because, for a sufficiently strong coupling $J\gg \omega_0$, thus
$\alpha_1=\sin\theta_\ua\approx 0$,  the off-diagonal terms in Eq. (\ref{H1})
which induce the oscillation between
two states with $\rho=1$ are vanishing and thus the occupation probabilities
of  the  $|10\rangle$ and $|11\rangle$ states are not changed.
As a result, the states $|10\rangle$ and $|11\rangle$ preserve their initial occupation probabilities,
while the states $|00\rangle$ and $|01\rangle$ experience a Rabi-type oscillation.
In this limit, the CNOT gate operation can be achieved with arbitrary parameter values.

In Fig. \ref{PC} we show the Rabi-type oscillation for strongly coupled qubits.
While the $P_{00}$ ($P_{01}$) is reversed  from 0.5 (0) to 0 (0.5)
at $\Omega t=$ (odd) $\pi$, we can observe that the probabilities $P_{10}$ and $P_{11}$
remain their initial values 0.5 and 0, respectively.
In this case the parameters need not satisfy the commensurate condition of Eq. (\ref{condition})
for the CNOT gate operation. 


\begin{figure}[b]
\includegraphics[height=5.5cm]{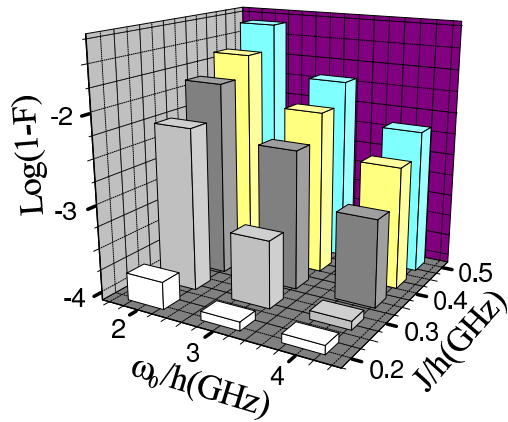}
\vspace*{-0.cm}
\caption{(Color online)
The errors of fidelity, $\delta F=1-F$, at the main peak for various values of $J$ and $\omega_0$
at $\Omega_0 t=\pi$.}
\label{dF}
\end{figure}

In fact, however, it is not so easy to obtain a sufficiently strong coupling
between qubits in experiment. Instead, we can achieve a high fidelity of CNOT
gate by choosing parameters satisfying Eq. (\ref{g}).
In real experiments  we can control the amplitude of the oscillating field $g$,
while the coupling $J$ and  energy gap $\omega_0$ are fixed.
In Table \ref{table} we compare the numerical results for $g$ at
the main fidelity maximum points ($n$=1) with those obtained from Eq. (\ref{g}) in the RWA,
which shows that two values fit well with each other. The
approximate values are a little smaller than the numerical values
and the small deviation tends to increase as $J$ increases and
$\omega_0$ decreases.
This can be understood from Eq. (\ref{ga}) where $g$ increases as
$J$ increases and $\omega_0$ decreases. Hence for small $J$ and
large $\omega_0$ the RWA and the numerical calculation coincide
with each other, because the RWA works well in the regime $g/\hbar
\ll \omega_0$. For the parameters far away from this regime the
two-qubit oscillation deviates seriously from the Rabi oscillation
and thus the CNOT gate operation cannot be performed, except the
strong coupling limit discussed.

\begin{table}[t]
\begin{center}
\vspace{0.5cm}
\begin{tabular}{c |c c c c c c}
\hline
\hline
 $J/h$ &                       & ~0.1 & ~0.3 & ~0.5  & ~0.7 & ~0.9     \\
\hline
$g/h$            & numerical  &  ~0.011 &  ~0.100 & ~0.265 & ~0.489 & ~0.754 \\
($\omega_0/2\pi=$4) & RWA    &  ~0.011 &  ~0.100 & ~0.264 & ~0.484 & ~0.744\\
\hline
$g/h$       & numerical      &  ~0.023 &  ~0.185 & ~0.448 & $\times$ & $\times$ \\
($\omega_0/2\pi=$2) & RWA    &  ~0.023 &  ~0.184 & ~0.443 & ~0.752 & ~1.090 \\
\hline \hline
\end{tabular}
\end{center}
\vspace{-0.cm}
\caption{ The values of $g/h$ for the main fidelity maxima ($n=1$) obtained from numerical
calculation and from the RWA of Eq. (\ref{g})   for
various coupling $J$ and qubit energy gap  $\omega_0$. For small $\omega_0$
and large $J$ the oscillations are far from the Rabi oscillation.
Here, the unit of all numbers is GHz.}
\label{table}
\end{table}


Though the value of fidelity at the main peak in Fig. \ref{peak} is close to the
maximum value of 1, it has small deviation, $\delta F =1-F$. In
Fig. \ref{dF} we plot the fidelity error $\delta F$ for various
values of $(\omega_0, J)$ at $\Omega_0 t = \pi$.
For large $\omega_0$ and small $J$ the fidelity error is vanishing,
and the Rabi oscillation by the Hamiltonian ${\tilde{\cal H}}$
in Eqs. (\ref{tilH0}) and (\ref{tilH1}) is
close to $\delta F \approx 10^{-4}$  for the fault-tolerant quantum
computing.
Hence for a weak coupling as well as a strong coupling
we can  achieve high performance  CNOT gate operation.

\section{Summary}

The commensurate oscillations of resonant and non-resonant modes
enable the high fidelity CNOT gate operation by finely tuning  the oscillating field  amplitude
for any given values of qubit energy gap and coupling strength between qubits.
While for a sufficiently strong coupling
the  CNOT gate can be achieved for any given parameter values,
for a weak coupling  a relation between the parameters should be satisfied
for the fidelity maxima.
For a sufficiently weak coupling compared to the qubit energy gap, $J/\hbar\omega_0\ll 1$,
we have $\alpha_1\approx 1$ and $\beta_1\approx 0$,  resulting in the
expression for $g$ in Eq. (\ref{ga}).
For $J/\hbar\omega_0\ll 1$, Eq. (\ref{ga}) immediately gives rise to the relation $g/J\ll 1$
and thus $g/\hbar\omega_0\ll 1$ after some manipulation.
This means that for a weak coupling $J/\hbar\omega_0\ll 1$ the numerical results are well fit with
the RWA as shown in Table \ref{table}, because the RWA is good for $g/\hbar\omega_0\ll 1$.
As a result, the high performance  CNOT gate operation
can be achieved  as shown in Fig. \ref{dF}.

\end{document}